\documentstyle[12pt]{article}

\begin{document}
\title{Wakes in Dilatonic Current-Carrying Cosmic Strings}
\author{A. L. N. Oliveira$^{1}$ and M. E. X. Guimar\~aes$^{2}$  \\
\mbox{\small{1. Instituto de F\'{\i}sica, Universidade de Bras\'{\i}lia }} \\
\mbox{\small{2. Instituto de F\'{\i}sica Te\'orica/UNESP}} \\
\mbox{\small{\bf  andreo@fis.unb.br, emilia@ift.unesp.br}}}
\maketitle
\date{}
\begin{abstract}
In this work, we present the gravitational field generated by a cosmic string
carrying a timelike current in the scalar-tensor gravities. The
mechanism of formation and evolution of wakes is fully investigated in
this framework. We show explicitly that the inclusion of electromagnetic
properties for the string induces logarithmic divergences in the accretion
problem.
\end{abstract}

\section{Introduction:}

Topological defects are predicted in many gauge models as solitonic
solutions resulting from spontaneous breaking of gauge or global symmetries.
Among all these solutions, cosmic strings have attracted attention because
they may be the source of large-scale structure in the Universe \cite{vil}. A
relevant mechanism to understand the structure formation by cosmic strings
involves long strings moving with relativistic speed in the normal plane,
giving rise to velocity perturbations in their wake \cite{silk}. The
mechanism of forming wakes has been considered by many authors in the
General Relativity theory \cite{vacha1,vacha2}. More recently, Masalskiene
and Guimar\~aes \cite{sandra} have considered the  formation of wakes by
long strings in the contexte of scalar-tensor theories. They have shown that
the presence of a gravitational scalar field - which from now on we will call
generically as ``dilaton" -  induces a very similar structure as in the case
of a wiggly cosmic string \cite{vacha2}. Further, Bezerra and Ferreira
\cite{bezerra} showed that, if a torsion is presented, this effect is
amplified.

Our purpose in this paper is twofold. We first generalize the works of Peter and Puy \cite{puy} and Ferreira et al. \cite{cris} for the case of
a cosmic string carrying a timelike current in the framework of scalar-tensor gravities. From the point of view of purely gravitational physics, it was shown that the inclusion of a current hardly affects the metric outside the string \cite{peter2}. However, it is well-known that inclusion of such an internal structure can change the predictions of cosmic string models in the microwave background anisotropies \cite{peter3,peter4}. This is our second goal. Namely, we analyse the  formation and evolution of wakes in this spacetime and we show explicitly how the current affects this mechanism. For this purpose, we consider a model in which non-baryonic cold dark matter propagates around the current-carrying string. The Zel'dovich approximation is carried out in order to treat this motion. We anticipate that our main result is to show that the inclusion of a current brings logarithmic divergences and can actually break down the accretion mechanism by wakes.

This work is outlined as follows. In section 2, after setting the relevant microscopic model which describes a superconducting string carrying a current of timelike type, we present its  gravitational field and we show the explicit dependence of the deficit angle, the deflection of light and geodesics on the timelike current. In section 3, we consider the mechanism of formation and evolution of wakes in this framework by means of the Zel'dovich approximation. Finally, in section 4, we end up with some conclusions and remarks.

\section{Timelike Current-Carrying Strings:}

In this section we will study the gravitational field generated by a string carrying a current
of timelike-type. We start with the action in the Jordan-Fierz frame
\begin{equation}
\label{acao_JF}
 {\cal{S}} = \frac{1}{16 \pi} \int d^4 x \sqrt{-\tilde{g}} [ \tilde{\Phi} \tilde{R} - \frac{\omega(\tilde{\Phi})}{\tilde{\Phi}}
\tilde{g} ^{\mu \nu} \partial _{\mu} \tilde{\Phi} \partial _{\nu} \tilde{\Phi} ] + {\cal{S}} _{m} [\psi _m, \tilde{g} _{\mu \nu}]
\end{equation}
$\tilde{g}_{\mu\nu}$ is the physical metric which contains both scalar
and tensor degrees of freedom, $\tilde{R}$ is the curvature scalar
associated to it and ${\cal S}_{m}$ is the action for general matter
fields which, at this point, is left arbitrary. The metric signature
is assumed to be $(-,+,+,+)$.

In what follows, we will concentrate our attention to superconducting vortex configurations which arise from the spontaneous breaking of the symmetry $U(1) \times U_{em}(1)$. Therefore,
the action for the matter fields will be composed by two pairs of coupled complex scalar and gauge fields $(\varphi, B_{\mu})$ and $(\sigma, A_{\mu})$.
Also, for technical purposes, it is preferable to work in the so-called Einstein (or conformal) frame, in which the scalar and tensor degrees of freedom do not mix.
\begin{eqnarray}
{\cal{S}}&=& \frac{1}{16 \pi G_*} \int d^4x \sqrt{-g} [R-2\partial _{\mu} \phi  \partial^{\mu} \phi ] \nonumber \\
&+& \int d^4x \sqrt{-g} [ -\frac{1}{2}\Omega ^2 (\phi) ((D_{\mu}\varphi) ^{*} D^{\mu} \varphi + (D_{\mu} \sigma) ^{*} D^{\mu} \sigma) \nonumber \\
&-& \frac{1}{16 \pi}  ( F_{\mu \nu} F^{\mu \nu} + H_{\mu \nu} H^{\mu \nu}) - \Omega ^2(\phi)  V(|\varphi|,|\sigma|) ] ,
\end{eqnarray}
where
$F_{\mu \nu} = \partial{\mu} A_{\nu} - \partial _{\nu} A_{\mu}$ ,
$H_{\mu \nu }= \partial{\mu} B_{\nu} - \partial _{\nu} B_{\mu}$
and the potential is suitably chosen in order that the pair $(\varphi, B_{\mu})$ breaks one symmetry $U(1)$ in vacuum (giving rise to the vortex configuration) and the second pair $(\sigma, A_{\mu})$ breaks the symmetry $U_{em}(1)$ in the core of the vortex (giving rise to the superconducting properties)
\begin{equation}
V(|\varphi|,|\sigma|)= \frac{\lambda _{\varphi}}{8} (|\varphi|^2- \eta ^2)^2 + f(|\varphi|^2 - \eta ^2) |\sigma|^2 + \frac{\lambda _{\sigma}}{4}
|\sigma|^4 + \frac{m_{\sigma}^2}{2} |\sigma|^2
\end{equation}

We restrict, then, ourselves to the configurations corresponding to an isolated, static current-carrying vortex lying in the $z-axis$. In a cylindrical coordinate system $(t,r,z,\theta)$ such
that $r \geq 0$ and $0 \leq \theta < 2\pi$, we make the following ansatz:
\begin{equation}
\varphi = \varphi(r)e^{i\theta} \;\;\;\;\; B_{\mu}= \frac{1}{q}[Q(r)-1]
\end{equation}
The pair $(\sigma, A_{\mu})$, which  is responsible for the superconducting properties of the vortex,  is set in the form
\begin{equation}
\sigma= \sigma(r)e^{i\chi(t)} \;\;\;\;\; A_t = \frac{1}{e}[P_t(r)-\partial_t
\chi]
\end{equation}
where $P_t$ corresponds to the electric field which leads to a timelike current in the vortex. We also require that the functions $\varphi, Q(r), \sigma(r) \mbox{and} P_t$ must be regular everywhere and must satisfy the usual boundary conditions of vortex \cite{ole} and superconducting configurations \cite{witten,carter}.

The action (2) is obtained from (\ref{acao_JF}) by a conformal transformation
\begin{eqnarray*}
\tilde{g} _{\mu \nu} = \Omega ^2(\phi) g_{\mu \nu}
\end{eqnarray*}
and by the redefinition of the quantity
\begin{eqnarray*}
G_* \Omega ^2(\phi)= \tilde{\Phi} ^{-1}
\end{eqnarray*}
which makes evident the feature that any gravitational phenomena will be affected by the variation
of the gravitation constant $G_*$ in the scalar-tensor gravity, and by introducing a new parameter
\begin{eqnarray*}
\alpha ^2 = ( \frac{\partial \ln\Omega(\phi)}{\partial \phi} ) ^2=
 [ 2 \omega (\tilde{\Phi}) +3 ] ^{-1}
\end{eqnarray*}

Variation of the  action (2) with respect to the metric $g_{\mu\nu}$ and to the dilaton field $\phi$ gives the modified Einstein's equations  and a wave equation for the dilaton, respectively. Namely,
\begin{eqnarray}
\label{EE0}
G_{\mu \nu} &=& 2 \partial _{\mu} \phi \partial _{\nu} \phi -g_{\mu \nu}
g^{\alpha \beta} \partial _{\alpha}
\phi \partial _{\beta} \phi +8 \pi G_*T_{\mu \nu} \nonumber \\
\Box _{g} \phi &=& -4\pi G_*\alpha(\phi)T
\end{eqnarray}

Where $T_{\mu\nu}$ is the energy-momentum tensor which is obtained by
\begin{equation}
\label{E-M}
T_{\mu \nu} = \frac{-2}{\sqrt{-g}} \frac{\delta S_{mat}}{\delta g_{\mu\nu}} .
\end{equation}
We note, in passing, that, in the conformal frame, this tensor is not conserved providing us with an additional equation
\[
\nabla_{\mu} T ^{\mu}_{\nu}=\alpha (\phi)T\nabla _{\nu} \phi .
\]

In what follows, we will write the general static metric with cylindrical symmetry corresponding to the electric case in the form
\begin{equation}
\label{metrica1}
ds^2 = - e^{2\psi}dt^2 +
 e^{2(\gamma - \psi)}(dr^2 + dz^2) + \beta^2 e^{-2\psi}d\theta^2
\end{equation}
where $\psi, \gamma, \beta$ are functions of $r$ only.


The non-vanishing components of the energy-momentum tensor using (\ref{E-M}) are
\begin{eqnarray}
T^t_t &=& -\frac{1}{2} \Omega ^2(\phi) \left\{ -g^{tt} \sigma ^2 {P_t}^2 + g^{rr} \left[{\varphi ^\prime}^2 +{\sigma ^\prime}^2\right]+ g^{\theta \theta}
\varphi ^2Q^2 \right\} \nonumber \\
&-& \frac{1}{8\pi}  g^{rr} \left\{-g^{tt}
\frac{{P_t^\prime}^2}{e^2} + g^{\theta \theta} \frac{{Q^\prime}^2}{q^2} \right\} - \Omega ^4(\phi) V(\sigma,\varphi) \nonumber \\
T^r_r&=& -\frac{1}{2} \Omega ^2(\phi) \left\{ g^{tt} \sigma ^2 {P_t}^2 - g^{rr} \left[{\varphi ^\prime}^2 +{\sigma ^\prime}^2\right]+ g^{\theta \theta}
\varphi ^2Q^2\right\} \nonumber \\
&+& \frac{1}{8\pi}  g^{rr} \left\{ g^{tt}
\frac{{P_t^\prime}^2}{e^2} + g^{\theta \theta} \frac{{Q^\prime}^2}{q^2} \right\} - \Omega ^4(\phi) V(\sigma,\varphi) \nonumber \\
T^{\theta}_{\theta}&=& -\frac{1}{2} \Omega ^2(\phi) \left\{ g^{tt} \sigma ^2 {P_t}^2 + g^{rr} \left[{\varphi ^\prime}^2 +{\sigma ^\prime}^2\right]- g^{\theta \theta}
\varphi ^2Q^2 \right\} \nonumber \\
&-& \frac{1}{8\pi}  g^{rr} \left\{ g^{tt}
\frac{{P_t^\prime}^2}{e^2} - g^{\theta \theta} \frac{{Q^\prime}^2}{q^2} \right\} - \Omega ^4(\phi) V(\sigma,\varphi) \\
T^z_z&=& -\frac{1}{2} \Omega ^2(\phi) \left\{ g^{tt} \sigma ^2 {P_t}^2 + g^{rr} \left[{\varphi^\prime}^2 +{\sigma ^\prime}^2\right]+ g^{\theta \theta}
\varphi ^2Q^2 \right\} \nonumber \\
&-& \frac{1}{8\pi}  g^{rr} \left\{ g^{tt}
\frac{{P_t^\prime}^2}{e^2} + g^{\theta \theta} \frac{{Q^\prime}^2}{q^2} \right\} - \Omega ^4(\phi) V(\sigma,\varphi). \nonumber
\end{eqnarray}

Therefore, for the electric case, eqs. (6) are written as
\begin{eqnarray}
\label{EE1}
\beta ^{\prime \prime} &=& 8 \pi G_{*} e^{2(\gamma -\psi)} \beta [T_1^1+T_3^3] \nonumber \\
(\beta \psi ^{\prime})^{\prime} &=& 4 \pi G_{*} e^{2(\gamma - \psi)} \beta [-T_0^0+T_1^1+T_2^2+T_3^3] \nonumber \\
\beta ^{\prime} \gamma ^{\prime}&=& \beta (\psi ^{\prime})^2 + \beta (\phi ^{\prime})^2 + 8 \pi G_{*} \beta e^{2(\gamma - \psi)}T_1^1 \nonumber \\
(\beta \phi ^{\prime})^{\prime} &=& -4 \pi G_{*} \beta e^{2(\gamma -\psi)}T
\end{eqnarray}

In order to solve the above equations we will divide the space in two regions: the exterior region, $r \geq r_0$, in which only the electric component of the Maxwell tensor contributes to the energy-momentum tensor and the internal region, $0 \leq r < r_0$, where all matter fields survive. $r_0$ is the string thickness.

\subsection{The Exterior Metric:}

Due to the specific properties of the Maxwell tensor
\begin{equation}
T^{\mu}_{\mu} =0 \hspace{1cm} \mbox{and} \hspace{1cm} T^{\alpha}_{\nu}T^{\mu}_{\alpha} = \frac{1}{4} (T^{\alpha \beta} T_{\alpha \beta}) \delta^{\mu}_{\nu}
\end{equation}
the Einstein's equations (\ref{EE1}) may be transformed into some algebraic
relations called Rainich algebra \cite{louis,melvin} which, for the  electric case, have the form\footnote{In the scalar-tensor theories, these relations are modified by a term which depend on the dilaton \cite{cris}.}
\begin{eqnarray}
R&=&R^{z}_{z}+R^{r}_{r}+R^{\theta}_{\theta}+R^{t}_{t}=2e^{2(\psi - \gamma)} {\phi ^\prime} ^2 \nonumber \\
(R^{t}_{t})^2 &=& (R^{r}_{r}-2g^{rr} {\phi ^\prime}^2)^2 = (R^{\theta}_{\theta})^2=(R^z_z)^2 \nonumber \\
R^t_t&=&-R^{\theta}_{\theta} \hspace{0.5cm} R^t_t=R^r_r-2g^{rr} {\phi ^\prime}^2 \hspace {0.5cm} R^{\theta}_{\theta}=R^z_z
\end{eqnarray}
Then we have
\begin{equation}
\psi ^{\prime \prime}+\frac{1}{r} \psi ^{\prime} - {\psi ^{\prime}}^2 - \gamma ^{\prime \prime} = \frac{D^2}{r^2}
\end{equation}
\begin{equation}
\gamma ^{\prime \prime} + \frac{1}{r} \gamma ^{\prime} =0
\end{equation}
the solution is
\begin{eqnarray}
\gamma &=& m^2 \ln\left(\frac{r}{r_0}\right) \nonumber \\
\psi &=& n \hspace{0.1cm} \ln\left(\frac{r}{r_0}\right) - \ln\left(\frac{\frac{r}{r_0}+k}{1+k}\right)
\end{eqnarray}

Therefore, the exterior metric for a timelike current-carrying string is:
\begin{eqnarray}
\label{extele}
ds^2_E&=&\left(\frac{r}{r_0}\right)^{2m^2-2n}W^2(r)(dr^2+dz^2)+ \left(\frac{r}{r_0}\right)^{-2n}W^2(r)
B^2r^2d\theta^2\nonumber\\
&-&\left(\frac{r}{r_0}\right)^{2n}\frac{1}{W^2(r)}dt^2
\end{eqnarray}
where
\begin{equation}
\label{w}
W(r)\equiv \frac{\left(\frac{r}{r_0}\right)^{2n}+k}{1+k} \, .
\end{equation}
The constants $m , n, k, D$ will be determined after the inclusion of the matter fields.

\subsection{The Internal Metric and the Matching:}

In order to solve the Einstein's equations (\ref{EE0}) in the internal region (where all matter fields will contribute to the energy-momentum tensor) we will
use the weak-field approximation. Therefore, we can expand the metric and the dilaton fields to first order in $G_0$ such that
\begin{eqnarray*}
g_{\mu\nu} & =  & \eta_{\mu\nu} + h_{\mu\nu} \\
\phi & = & \phi_0 + \phi_1
\end{eqnarray*}
In this way, eqs. (\ref{EE0}) reduce to
 \begin{equation}
  \label{linear1}
  \nabla^2h_{\mu\nu}=-16\pi G_0(T^{(0)}_{\mu\nu}-\frac{1}{2}\eta_{\mu\nu}T^{(0)}).
  \end{equation}
in a harmonic coordinate system such that $(h_{\nu}^{\mu}-\frac{1}{2}\delta^{\mu}_{\nu}h)_{,\nu}=0$, and the linearised equation for the dilaton field
 \begin{equation}
   \label{linear2}
   \nabla^2\phi_{(1)}= -4\pi G_0 \alpha(\phi_0)T^{(0)}.
   \end{equation}
In the above eqs. (\ref{linear1}) and (\ref{linear2}), $T^{(0)}_{\mu\nu}$ is the energy-momentum tensor evaluated to zeroth-order in $G_0$, and $T^{(0)}$ its trace. In cartesian (harmonic) coodinates the nonvanishing componentes of $T^{(0)}_{\mu\nu}$ are
\begin{eqnarray}
\label{EM0}
T_{(0)t}^t &=& -\frac{1}{2} \left\{  \sigma ^2 {P_t}^2 +  {\varphi ^\prime}^2 +{\sigma ^\prime}^2+ \frac{1}{r^2}
\varphi ^2Q^2  \right\} \nonumber \\
&-& \frac{1}{8\pi}  \left\{
-\frac{{P_t^\prime}^2}{e^2} + \frac{1}{r^2} \frac{{Q^\prime}^2}{q^2} + \right\} -  V(\sigma,\varphi) \nonumber\\
T_{(0)x} ^x &=& \left(\cos^2(\theta) -\frac{1}{2}\right) \left\{{\sigma ^\prime}^2+ {\varphi ^\prime}^2 - \frac{\varphi ^2 Q^2}{r^2}
-\frac{{P_t^\prime}^2}{4\pi e^2}\right\} \nonumber\\
&-&\frac{1}{2}\left\{- \sigma^2 P_t^2 - \frac{1}{4\pi}  \frac{Q^\prime}{r^2q^2} + 2  V(\sigma,\phi)\right\} \nonumber\\
T_{(0)y} ^y &=& \left(\sin^2(\theta)-\frac{1}{2}\right) \left\{ {\varphi ^\prime}^2+{\sigma ^\prime}^2 - \frac{\varphi ^2Q^2}{r^2}
-\frac{{P_t^\prime}^2}{e4 \pi ^2}\right\} \nonumber \\
&-& \frac{1}{2} \left\{-\sigma ^2 P_t ^2 -
\frac{1}{4\pi}  \frac{{Q^\prime}^2}{r^2q^2} + 2  V(\sigma,\phi)\right\} \nonumber\\
T_{(0)z} ^z &=& -\frac{1}{2} \left\{ - \sigma ^2 {P_t}^2 +  {\varphi ^\prime}^2 +{\sigma ^\prime}^2+ \frac{1}{r^2}
\varphi ^2Q^2 \right\} \nonumber \\ &-& \frac{1}{8\pi}   \left\{ -
\frac{{P_t^\prime}^2}{e^2} + \frac{1}{r^2} \frac{{Q^\prime}^2}{q^2} \right\} -  V(\sigma,\varphi)
\end{eqnarray}
In this paper, in order to solve eqs. (\ref{linear1}) and (\ref{linear2}) with source given by (\ref{EM0}), we will use a method which was first applied by
Linet in ref. \cite{linet} and later by Peter and Puy \cite{puy} and Ferreira et al. \cite{cris}. The trick consists in writing down the string and its components fields by means of $\delta$ functions\footnote{The calculations are straightforward but lengthy. We will skip here the details and provide directly the results.
For the details of these calculations, we refer the reader to refs. \cite{puy,cris} .}.
In doing this, we are allowed to write the string's energy-momentum tensor as
\begin{equation}
 \label{e_m_linet}
 T^{\mu\nu}=diag \, (U,-Z,-Z,- T)
 \end{equation}
where $U, \; Z \; \mbox{and} \; T$ are macroscopic quantities which are defined as the energy per unit length, the transverse components per unit length and the tension per unit length, respectively:
\begin{eqnarray}
U & \equiv &  M^t_t  =  -2\pi\int_0^{r_0}T^t_t \; r \; dr  \nonumber \\
X & \equiv & M^r_r=-2\pi\int_0^{r_0}T^r_r \; r \; dr  \\
Y & \equiv  & M^{\theta}_{\theta}=-2\pi\int_0^{r_0}T_{\theta}^{\theta} \; r \; dr  \nonumber \\
T & \equiv  & M^z_z=-2\pi\int_0^{r_0}T^z_z \; r \; dr   \nonumber
\end{eqnarray}
From the transverse quantities $X$ and $Y$ we define one single cartesian component   $Z$:
\begin{equation}
Z=-\int r dr d\theta T_x^x=-\int r dr d\theta T_y^y
\end{equation}

On the other hand, the energy-momentum in the exterior region (and corresponding
definition of the electric current)
can be re-written in cartesian coordinates as
\begin{eqnarray}
\label{eletroele}
T^{tt}_{em}&=&T^{zz}_{em}=\frac{I^2}{2\pi r}, \nonumber\\
T^{kj}_{em}&=& - \frac{I^2}{2\pi r^4} (2 x^{k}\; x^j-r^2 \delta _{kj})
\end{eqnarray}
Therefore, we have
\begin{eqnarray}
T^{tt}_{(0)}&=&U \delta(x)\delta(y)+\frac{I^2}{4\pi} \nabla^2\left[\ln \left(\frac{r}{r_0}\right)\right]^2, \nonumber \\
T^{zz}_{(0)}&=& -T\delta(x)\delta(y)+\frac{I^2}{4\pi}\nabla^2\left[\ln\left(\frac{r}{r_0}\right)\right]^2, \\
T^{kj}_{(0)}&=& -  I^2\left[\delta^{kj}\delta(x)\delta(y)- \frac{\partial_k\partial_j\ln\left(\frac{r}{r_0}\right)}{2\pi}\right] \nonumber
\end{eqnarray}
Solving the linearised Einstein's equations, we obtain
\begin{eqnarray}
&&\nabla^2h_{tt}=-4G_0\left[I^2\nabla^2\left(\ln\left(\frac{r}{r_0}\right)\right)^2+ 2\pi(U-T -  I^2)\delta(x)\delta(y)\right] \nonumber\\
&\Rightarrow &h_{tt}=-4G_0\left[I^2\left(\ln\left(\frac{r}{r_0}\right)\right)^2+ (U-T -  I^2)\ln\left(\frac{r}{r_0}\right)\right]
\end{eqnarray}
\begin{eqnarray}
&&\nabla ^2h_{zz}=-4G_0\left[I^2\nabla^2 \left(\ln\left(\frac{r}{r_0}\right)\right)^2+ 2\pi(U-T +  I^2)\delta(x)\delta(y)\right]\nonumber\\
&\Rightarrow &h_{zz}=-4G_0\left[I^2\left(\ln\left(\frac{r}{r_0}\right)\right)^2+(U-T +  I^2)\ln\left(\frac{r}{r_0}\right)\right]
\end{eqnarray}
\begin{eqnarray}
&&\nabla^2h_{kj}=-4G_0\left[+2I^2\partial_k\partial_j\ln\left(\frac{r}{r_0}\right)+ 2\pi\delta_{kj}(U-T -  I^2)
\delta(x)\delta(y)\right] \nonumber \\
&\Rightarrow &h_{kj}= 2G_0I^2r^2\partial_k\partial_j\ln\left(\frac{r}{r_0}\right) - 4G_0\delta_{kj}(U+T - I^2)\ln\left(\frac{r}{r_0}\right)
\end{eqnarray}
In order to match the internal and the external solutions and obtain the parameters $m$, $B$, $n$ e $D$ we need to return to the cylindrical coordinates system
 \begin{eqnarray}
\label{sol}
 g_{tt}&=&-\left\{1+4G_0\left[I^2\left(\ln\left(\frac{r}{r_0}\right)\right)^2+(U-T - I^2)\ln\left(\frac{r}{r_0}\right)\right]\right\},\\
 g_{zz}&=&1-4G_0\left[I^2\left(\ln\left(\frac{r}{r_0}\right)\right)^2+(U-T +  I^2)\ln\left(\frac{r}{r_0}\right)\right] \\
g_{rr}&=&1- 2G_0 I^2-4G_0(U+T - I^2)\ln\left(\frac{r}{r_0}\right), \\
g_{\theta\theta}&=&r^2\left[1+2G_0 I^2-4G_0(U+T - I^2)\ln\left(\frac{r}{r_0}\right)\right].
\end{eqnarray}
Before presenting the final result we note that the metric (\ref{sol}) does not preserve the feature $g_{RR}=g_{zz}$. We need again to make a change of coordinates, to first order in $G_0$
 \[
 R=r\left[1+a_1-a_2\ln\left(\frac{r}{r_0}\right)-
 a_3\left(\ln\left(\frac{r}{r_0}\right)\right)^2\right],
 \]
where $a_1=G_0(4T-I^2)$, $a_2=4G_0T$ and $a_3=-2G_0I^2$. Without loss of generalities, let us make $R=r$ and expand the parameters in powers of $\ln\left(\frac{r}{r_0}\right)$. In doing that, we finally get
\begin{eqnarray}
\label{ele}
ds^2&=& \left\{1-4G_0\left[ \left(U-T+I^2\right)\hspace{0.1cm}\ln\left(\frac{r}{r_0}\right)+I^2\hspace{0.1cm}
\ln^2\left(\frac{r}{r_0}\right) \right]\right\} \left(dr^2+ dz^2\right) \nonumber \\
&+& r^2\left[ 1-8G_0\left(T-\frac{I^2}{2}\right) - 4 G_0 \left(U-T-I^2\right) \hspace{0.1cm} \ln \left(\frac{r}{r_0}\right) - 4 G_0 I^2 \ln ^2 \left(\frac{r}{r_0}\right) \right] d \theta ^2 \nonumber \\
&-& \left\{ 1+ 4G_0 \left[ I^2\hspace{0.1cm} \ln^2 \left(\frac{r}{r_0}\right) + \left(U-T-I^2\right) \hspace{0.1cm} \ln \left(\frac{r}{r_0} \right)\right] \right\} dt^2
\end{eqnarray}

In the next section, we will use this metric to compute the deficit angle and the deflection of light rays by a current-carrying string.

\subsection{Deficit Angle and the Deflection of Light Rays by a Current-Carrying String:}

Now that we have calculated the metric (\ref{ele}) of a timelike current-carrying string, we can compute some quantities associated to it, such as the deficit angle and the deflection of light rays.
The deficit angle is calculated with the help of the expression
\[
\delta _{\theta} = 2 \pi (1-\frac{1}{\sqrt{g _{\rho\rho}}} \frac{d}{d \rho}
 \sqrt{g_{\theta \theta}}) \, .
\]
For the metric (\ref{ele}) it is, to first order in $G_0$:
\begin{equation}
\label{deficit}
 \delta_{\theta}=4 \pi G_0\left( U + T - 2I^2 \right)+ O(G_0^2)
\end{equation}
It is clear that, if $I$ is zero (the neutral string case), the deficit angle
(\ref{deficit}) reduces to $\delta_{\theta} = 8\pi G_0 U$ which is of order
$\sim 10^{-6}$ for GUT strings.

The deflection of the light ray is given by the expression
\begin{equation}
\label{lightray}
\bigtriangleup \theta = 2 \int _{\rho _{min}}^{\infty}
d \rho [ - \frac{g^2_{\theta \theta} p^{-2}}
{g _{\rho \rho} g_{tt}} - \frac{g_{\theta\theta}}{g_{\rho \rho}} ]^{-1/2} -
 \pi
\end{equation}
As in \cite{puy} we obtain, knowing that $p$ is the impact parameter,
\begin{equation}
\label{deflection}
\Delta\theta = 4 \pi G_0 \left[ U- I^2 \left( \frac{3}{2} -
\ln\left(\frac{p}{r_0}\right) \right) \right] + 4 \pi \hspace{0.1cm}
\ln (2) I^2
G_0 + O(G_0 ^2) \, .
\end{equation}
From expression (\ref{deflection}) we can observe some remarkable
features of the
deflection of light due to a timelike current-carrying string in the scalar-tensor
gravity.
First of all, there is no dependence on the tension in the deflection
of light rays. This feature is a reminescence of the neutral string. The
second feature is the dependence in the impact parameter $p$ due to the
presence of the current. This is due to the fact that the inclusion of an
electromagnetic field actually increases the total energy with the
distance to the string core and, as a consequence, the deflection of light
is more intense. Finally, we notice that the dilaton field amplifies this
effect since its inclusion also contributes to increase the total energy
of the system.

\section{Formation and Evolution of the Wakes and the Zel'dovich Approximation:}

A relevant mechanism to understand the structure formation by cosmic strings involves long strings moving with relativistic speed in the normal plane, giving rise to velocity perturbations in their wake \cite{silk}. Matter through which a long string moves acquires a boost in the direction of the surface swept out by the string. Matter moves toward this surface by gravitational attraction and, as a consequence,  a wake is formed behind the string.
In what follows, we will study the implications of a timelike current on the formation and evolution of a wake behind the string which generates the metric
(\ref{ele}). For this purpose, we will mimic this situation with a simple model consisting of cold dark matter composed by non-relativistic collisionless particles moving past a long string. In order to make a quantitative description of accretion onto wakes, we will use the Zel'dovich approximation, which consists in considering the Newtonian accretion problem in an expanding universe using the method of linear perturbations.

To start with, we first compute the velocity perturbation of massive particles moving past the string. If we consider that the string is moving with normal velocity $v_s$ through matter, the velocity perturbation can be calculated with the help of the gravitational force due to metric (\ref{ele}):
\begin{eqnarray}
\label{boost}
u & = &  8\pi G_0 U v_s\gamma \nonumber \\
&  + &   \frac{\pi G_0}{v_s\gamma}\left[2\alpha(\phi_0)^2\left(U+ T +I^2\right)-\left(U- T -I^2\right)-2\ln\left(\frac{r}{r_0}\right)I^2\right]
\end{eqnarray}
with $\gamma = (1-v_s^2)^{-1/2}$.
The first term in (\ref{boost}) is equivalent to the relative velocity of particles flowing past a string in general relativity. The other terms come as a consequence of the scalar-tensor coupling of the gravitational interaction and the superconducting properties of the string.

Let us suppose now that the wake was formed at $t_i > t_{eq}$. The physical trajectory of a dark particle can be written as
\begin{equation}
\label{traj}
h(\vec{x}, t) = a(t) [ \vec{x} + \psi(\vec{x}, t)]
\end{equation}
where $\vec{x}$ is the unperturbed comoving position of the particle and  $\psi(\vec{x}, t)$ is the comoving displacement developed as a consequence of the gravitational attraction induced by the wake on the particle. Suppose, for simplification, that the wake is perpendicular to the $x$-axis (assuming that $dz=0$ in the metric (\ref{ele}) and $ r = \sqrt{x^2 + y^2}$) in such a way that the only non-vanishing component of $\psi$ is $\psi_x$. Therefore, the equation of motion for a dark particle in the Newtonian limit is
\begin{equation}
\label{newton}
\ddot{h} =  - \nabla_h \Phi
\end{equation}
where the Newtonian potential $\Phi$ satisfies the Poisson equation
\begin{equation}
\label{poisson}
\nabla_h^2 \Phi = 4\pi G_0 \rho
\end{equation}
where $\rho(t)$ is the dark matter density in a cold dark matter universe. For a flat universe in the matter-dominated era, $a(t) \sim t^{2/3}$. Therefore, the linearised equation for $\psi_x$ is
\begin{equation}
\label{psi}
\ddot{\psi} + \frac{4}{3t}\dot{\psi} - \frac{2}{3t^2}\psi = 0
\end{equation}
with appropriated initial conditions: $\psi(t_i) = 0$ and $\dot\psi(t_i) = -u_i$. Eq. (\ref{psi}) is the Euler equation whose solution is easily found
\[
\psi(x,t) = \frac{3}{5}\left[\frac{u_i t_i^2}{t} - u_i t_i \left(\frac{t}{t_i}\right)^{2/3}\right]
\]
Calculating the comoving coordinate $x(t)$ using the fact that $\dot{h} = 0$ in the ``turn around"\footnote{The moment when the dark particle stops expanding with the Hubble flow and starts to collapse onto the wake.}, we get
\begin{equation}
\label{comoving}
x(t) = - \frac{6}{5} \left[ \frac{u_i t_i^2}{t} - u_i t_i \left(\frac{t}{t_i}\right)^{2/3}\right]
\end{equation}
With the help of (\ref{comoving}) we can compute both the thickness $d(t)$ and the surface density $\sigma(t)$ of the wake \cite{vil}. We have, then, respectively (to first order in $G_0$)
\begin{eqnarray}
\label{thick}
d(t) & \approx & \frac{12}{5} \left(\frac{t}{t_i}\right)^{1/3} \left\{ 8\pi G_0 U v_s\gamma  +    \frac{\pi G_0}{v_s\gamma} \left[ 2\alpha(\phi_0)^2\left(U+ T +I^2\right)-\left(U- T -I^2\right)-2\ln\left(\frac{r}{r_0}\right)I^2\right] \right\} \nonumber \\
\sigma(t) & \approx & \frac{2}{5} \frac{1}{v_s\gamma t} \left(\frac{t}{t_i}\right)^{1/3} \left[
8U (v_s\gamma)^2 + 2\alpha(\phi_0)^2 \left(U+T+I^2\right)- \left( U-T-I^2\right) - 2\ln\left(\frac{r}{r_0}\right)I^2  \right]
\end{eqnarray}
where we have used the fact that $\rho(t) = \frac{1}{6\pi G_0 t^2}$ for a flat universe in the matter-dominated era and that the wake was formed at $t_i \sim t_{eq}$. Clearly, from eq. (\ref{thick}) we see that the presence of the current makes the accretion mechanism by wakes to  diverge.

\section{Conclusion:}

Inclusion of a current in the internal structure of a cosmic string could drastically change the predictions of such models in the microwave background anisotropies. In particular, a current of a timelike-type could bring some divergences leading to some unbounded gravitational effects \cite{peter2}.
In this work we studied the effects of a timelike-current string in the mechanism of
wakes formation.
For this purpose, we first studied the gravitational properties of the spacetime generated by this string in the framework of scalar-tensor gravities. We analysed the dependence of the deficit angle and  the deflection of  light  on the current in this spacetime. Then, we carried out an investigation of the mechanism of formation and evolution of wakes in this framework,  showing the explicit contribution of the current to this effect.

Wakes produced by moving strings can provide an explanation for filamentary and sheetlike structures observed in the universe. A wake produced by the string in one Hubble time has the shape of a strip of width $\sim v_s t_i$. With the help of the surface density (\ref{thick}) we can compute the wake's linear mass density, say $\tilde{\mu}$,
\begin{equation}
\sigma(t) \approx  \frac{2}{5} \frac{1}{\gamma t} \left(\frac{t}{t_i}\right)^{1/3} \left[
8U (v_s\gamma)^2 + 2\alpha(\phi_0)^2 \left(U+T+I^2\right)- \left( U-T-I^2\right) - 2\ln\left(\frac{r}{r_0}\right)I^2  \right]
\end{equation}
If the string moves slower or if we extrapolate our results to earlier epochs, we see that the logarithmic term would bring divergences and the mechanism of forming wakes would break down.

\section*{Acknowledgements:}

The authors would like to thank CAPES for partial financial support in the contexte of the PROCAD program. A. L. N. O. would like to thank CAPES for a PhD grant. M. E. X. G. is on leave from Depto. de Matem\'atica, Universidade de
Bras\'{\i}lia.

\end{document}